\documentstyle[12pt]{article}
\input psbox.tex
\input epsf.tex
\begin{document}
\setcounter{page}{0}
\vskip0.5cm
\begin{center}

{\large\bf Quarkonium Formation Time \\
\vskip0.2cm
in a Model-Independent Approach }

\vskip1cm

D. Kharzeev$^a$ and R. L. Thews$^b$\\

\vskip0.8cm
{\it $ ^a)$ 
RIKEN-BNL Research Center\\
Brookhaven National Laboratory\\
Upton, NY 11973 USA\\
\medskip
$ ^b)$ Department of Physics\\
University of Arizona\\
Tucson, AZ 85721 USA}
\end{center}
\vskip1cm
\begin{abstract}
We use dispersion relations to reconstruct, in a model--independent 
way, the formation dynamics of heavy quarkonium from the experimental data 
on $e^+e^- \to \bar{Q}Q$ annihilation. We extract a distribution
of formation times  
with a mean value for the $J/\psi$, $ <\tau_{J/\psi}> = 0.44\ {\rm fm}$;
 and for the $\Upsilon$, $<\tau_{\Upsilon}> 
= 0.32\ {\rm fm}$. The corresponding widths of these distributions are
given by $\Delta\tau_{J/\psi} = 0.31\ {\rm fm}$ and 
$\Delta\tau_{\Upsilon} 
= 0.28\ {\rm fm}$. This information can be used as an input in modeling 
of heavy quarkonium production on nuclear targets.
\end{abstract}

\vspace {-0.10in}

\vskip5cm

\newpage

The creation of a heavy quark-antiquark pair occurs at small distances 
($\sim 1/m_Q$) and  produces compact
$\bar{Q}Q$ states which later transform into physical heavy hadrons.
In the case of quarkonium production on nuclear targets, this evolution 
can cause observable effects \cite{BM}. While these ``formation" \cite{EF} effects
can 
in principle be evaluated in Glauber--Gribov theory \cite{Gribov}, in practice 
this calculation is difficult to perform in a model--independent way, 
since it requires the knowledge of all off-diagonal components of 
the quarkonium--nucleon scattering amplitude. Therefore one often uses 
a simplistic approach, in which the evolution of the quark-antiquark 
pair is mimicked by a fixed ``formation time", during which the 
interactions of the pair are different from the interactions of the 
physical quarkonium; depending on the color state of the pair, the 
interactions of the ``unformed" pair can be either suppressed 
(``color transparency" \cite{BM}) or enhanced, if the pair are formed 
in the color octet state \cite{KS}.

In the literature one can find different prescriptions for the formation 
time $\tau_f$. A still popular viewpoint, for instance, is to assume  
a universal 
parameter on the order of some characteristic hadronic scale, say, $\tau_f\sim
m_{\rho}^{-1}$. Alternatively, one considers the classical expansion 
of the heavy quark pair and defines formation time as the time 
when the separation of the pair reaches the size of 
a physical quarkonium state 
\cite{GF}. Much work has been done also on the quantum-mechanical approach to 
quarkonium formation, where the expansion of a small initial wave packet
is controlled by the spacings of the bound state mass spectrum
\cite{RT},\cite{rev}.  
In this work we address the problem of formation time 
starting from the idea that all essential information about the expansion 
of the wave packet is contained in the correlator of the
hard scattering operator \cite{DK}.
  
Let us consider the space-time correlator of an operator $\hat J$, which 
produces from initial state $\mid i>$ the 
$(\bar{Q}Q)$ state with certain quantum numbers
\begin{equation}
\Pi(x)=\left<i\mid T\{\hat J(x)\hat J(0)\}\mid i\right> 
\label{1}
\end{equation} 
The basic expression which allows the use of experimental data for 
extracting 
 information from this correlator is the dispersion relation, 
which in coordinate representation \cite{BS}, \cite{ES} takes the form:
\begin{equation}
\Pi(x)=\frac{1}{\pi}\int Im\Pi(s) D\left(\sqrt s,x^2\right)ds, \label{2}
\end{equation}
where
\begin{equation}
 D\left(\sqrt s,\tau^2=-x^2\right)=\frac{\sqrt s}{4\pi^2\tau} 
K_1\left(\sqrt s \tau\right) \label{3}
\end{equation} 
is the relativistic causal propagator in the coordinate representation; 
$K_1$ is the Hankel function.
The expression (\ref{2}) relates the behavior of the correlator (\ref{1}) 
to experimentally measurable cross sections for physical processes. 
For example, in the case 
of $(\bar{Q}Q)$ pair production in $e^+e^-$ annihilation one has
simply
\begin{equation}
Im\Pi(s)={s \over (4\pi\alpha)^2}\ \sigma(e^+e^-\rightarrow \bar{Q}Q; s). 
\label{4}
\end{equation}

The physical meaning of eq.(\ref{2}) is transparent: it represents the 
correlator as a superposition of propagators of physical states, 
each with the weight proportional to
the probability of their production in a hard process.

Thus it is possible to extract information  
about the space-time evolution of various
states in a given hadronic channel with fixed 
quantum numbers directly from experimental data. For example, 
we can define the formation time of the ground state with mass m 
by the time $\tau_f$ at which 
the correlator 
approaches its asymptotic behavior
\begin{equation}
\Pi(\tau) \sim \tau^{-3/2}\ exp(-m\tau); \label{formt} 
\end{equation}
note that $\tau=it$ is Euclidian time. 

To illustrate the notion of formation 
time in more detail, let us use the following simple example -- assume that the
spectral 
density (\ref{4}) consists of two narrow states of identical strength, i.e.
\begin{equation}
Im\Pi(s) \sim \delta(s-m_1^2) +  \delta(s-m_2^2). \label{ex}
\end{equation} 
At large time $\tau$, the correlator (\ref{1}) will look like
\begin{equation}
\Pi(\tau) \sim \tau^{-3/2}\ exp(-m_1\tau) \left[ m_1^{1/2} + m_2^{1/2}
\  exp(-(m_2-m_1)\tau) \right]. 
\end{equation}
The use of criterion (\ref{formt}) therefore leads to 
\begin{equation}
\tau_f \sim \frac{1}{m_2-m_1}. \label{form12}
\end{equation} 
Let us now introduce invariant $\lambda \equiv t^2 - r^2$ and 
decompose (\ref{3}) in the following way \cite{BS}:
\begin{equation}
 D\left(\sqrt s, \lambda \right) = {\frac{1}{4\pi}}\ \delta(\sqrt{\lambda}) - 
{\frac{\sqrt s}{8 \pi \sqrt \lambda}}\ \theta(\lambda) 
\left[ J_1(\sqrt s \sqrt \lambda) - i N_1(\sqrt s \sqrt \lambda)\right] +
\label{dec}
\end{equation}
$$ 
+ {\frac{i \sqrt s}{4 \pi^2 \sqrt{-\lambda}}}\ \theta(-\lambda) K_1(\sqrt s
\sqrt{-\lambda});
$$
It is clear from (\ref{dec}) that the formation time $\tau_f$ extracted from 
the Euclidian 
asymptotics (\ref{formt}) of the correlator (\ref{2}) will also determine the
propagation 
of the quark-antiquark state in Minkowski space, with $\lambda >0$. Let us
introduce 
the light-cone variables $x^+ = t + z\ $, $x^- = t - z$, so that 
$\lambda =  t^2 - r^2 = x^+ x^-$, and conjugate momenta $p^- = p_0 - p_z \simeq
m^2/2p_z$ and 
$p^+ = p_0 + p_z$;
as can be seen from (\ref{2}), (\ref{ex}) and (\ref{dec}), in Minkowski space the 
criterion (\ref{formt}) 
leads to
\begin{equation}
l_f = \Delta x^+ \sim \frac{1}{\Delta p^-} \simeq \frac{2 p}{m_2^2 - m_1^2},
\label{forml} 
\end{equation}
i.e. the formation length of quarkonium $l_f$ grows linearly with its 
momentum $p=p_z$ in the lab frame. The formation time (\ref{form12}) extracted
from Euclidian 
asymptotics (\ref{formt})  
of the correlator is related to the formation length (\ref{forml}) by the Lorentz
transformation: 
\begin{equation}
l_f\simeq \frac{p}{(m_1+m_2)/2}\ \tau_f = \frac{2p}{m_2^2-m_1^2},\label{6}
\end{equation}
where $(m_1+m_2)/2$ can be interpreted as a characteristic mass of the wave
packet. 
One can recognize this length also as the inverse of the longitudinal
momentum transfer \cite{Gribov}
\begin{equation}
\Delta q_{\parallel} \simeq \sqrt{p_0^2 - m_1^2} - \sqrt{p_0^2 - m_2^2} \simeq
\frac{m_2^2-m_1^2}{2p}, 
\end{equation}
which accompanies the transition between the two hadronic states at high energy.
The wave packet can escape from a nucleus of radius $R_A$ before 
being "formed" if $l_f\gg R_A$ or, equivalently, if
\begin{equation}
\frac{m_2^2-m_1^2}{2p}R_A\ll 1. \label{7}
\end{equation}
This condition was derived many years ago by Gribov \cite{Gribov}.

One can clearly see that the value of formation time is by no means a 
universal constant. It depends both on the properties of the interaction 
$\hat J$ and on 
the entire spectrum of states in a 
given hadronic channel. Models using some 
universal value of formation time are therefore misleading.
Let us note that eqs.(1-4) 
indicate that 
the space-time picture of a hard process can be equivalently described 
in the language of the spectrum of hadronic excitations which 
are formed as a result of this hard process.
If the operator $J$ is of a short-range nature, then it produces a compact 
object;  Eq.(\ref{2}) tells us 
that it is composed of many normal-sized hadronic states. 

Since there exist high quality data on the production of 
heavy quark states 
in $e^+e^-$ annihilation, we will concentrate in the remainder of this Letter 
on the formation dynamics of vector $J^{PC}=1^{--}$ quarkonium states. 
In this case the Fourier transform of the correlator 
$\Pi_{\mu\nu} = \langle 0| T\{J_{\mu}(x)J_{\nu}(0)\} |0 \rangle$ has the 
following familiar structure:
\begin{equation}
i\int d^4x\ e^{iqx}\ \Pi_{\mu\nu}(x) = 
\Pi(q^2) (q_{\mu}q_{\nu} - q^2 g_{\mu\nu}). 
\label{9}
\end{equation}
We use the cross section for $e^+e^-\to
\mu^+\mu^-$ annihilation, 
\begin{equation}
\sigma(e^+e^-\to\mu^+\mu^-;s) = {4\pi\alpha^2 \over 3s},
\end{equation}
to express (\ref{4}) in terms of the familiar ratio 
\begin{equation}
R(s) = {  \sigma(e^+e^-\to\bar{Q}Q;s) \over \sigma(e^+e^-\to\mu^+\mu^-;s)},
\end{equation}
resulting in
\begin{equation}
Im \Pi(s) = {1 \over 12 \pi} R(s).
\end{equation}
Contracting Lorentz indices and using the relation 
$(q_{\mu}q_{\nu} - q^2 g_{\mu\nu})|_{\mu=\nu} = 3 q^2 = -3 \partial^2$, 
one can write down the expression
(\ref{2}) as \cite{BS}, \cite{ES}
\begin{equation}
\Pi(x) = -3 {1 \over 12\pi^2} \int_0^{\infty} R(s)\ \partial^2 
D(\sqrt{s}, x^2) ds .
\end{equation}
Since the scalar propagator satisfies the equation
\begin{equation}
(-\partial^2 - s) D(\sqrt{s}, x^2) = - \delta(x),
\end{equation}
at $x^2\neq 0$ the final result takes the following form:
\begin{equation}
\Pi(x) = {1 \over 4 \pi^2} \int_0^{\infty} s R(s) D(\sqrt{s}, x^2) ds . 
\label{15}
\end{equation}
We parametrize the ratio $R(s)$ in terms of narrow resonances plus
a continuum contribution:
\begin{equation}
R(s) = \sum_i R_i(s) + R_{cont}(s), \label{16}
\end{equation}
where 
\begin{equation}
R_i(s) = (2J_i+1) {3 \over 4 \alpha^2} {B(\Psi_i \to e^+e^-) \Gamma_i^2 
\over {(\sqrt{s} - M_i)^2 + {\Gamma_i^2 \over 4}}} \label{17}
\end{equation}
is the contribution of the $\Psi_i$ resonance with spin $J_i$, mass $M_i$ and 
total width $\Gamma_i$. The continuum contribution 
can be described with a reasonable accuracy 
as 
\begin{equation}
R_{cont}(s) = 3 e_Q^2 \Theta (s - s_{th}),
\end{equation} 
where $e_Q$ is the electric charge of the heavy quark.

>From (\ref{3}), (\ref{15}), and (\ref{17}) we calculate the contribution
of each narrow resonance term 
\begin{equation}
\Pi_i(\tau)={3 \sigma_i \Gamma_i {M_i}^6 \over 
64 \pi^4 \alpha^2 \tau} K_1 (M_i \tau),
\label{PIRES}
\end{equation}
where 
\begin{equation}
\sigma_i = {4 \pi \over {M_i}^2} (2 J_i + 1) B(\Psi_i \to e^+e^-)
\end{equation}
is the magnitude of the annihilation cross section at the
resonance peak $\sqrt{s} = M_i$.

The continuum contribution is

\begin{equation}
\Pi_{cont}(\tau) = { 3 {e_Q}^2 \over 8 \pi^4 \tau^6}
\int_{2 M_{th} \tau}^{\infty} x^4 K_1(x) dx, 
\label{PICONT}
\end{equation}
where the open flavor threshold $s_{th} = 4 {M_{th}}^2$.

The standard limiting forms of the Bessel function then give
the behavior of these terms for large and small Euclidean times:

\begin{equation}
\Pi_i(\tau \to \infty) =
{3 \sigma_i \Gamma_i {M_i}^{11 \over 2} \over 64 \pi^3 \alpha^2
\sqrt{2\pi} \tau^{3 \over 2}} e^{-M_i\tau}
\label {PIRESTINF}
\end{equation}
\begin{equation}
\Pi_{cont}(\tau \to \infty) =
{3 {e_Q}^2 {M_{th}}^{7 \over 2} \over \pi^{7 \over 2} 
 \tau^{5 \over 2}} e^{-2 M_{th}\tau}
\label {PICONTINF}
\end{equation}
\begin{equation}
\Pi_i(\tau \to 0) =
{3 \sigma_i \Gamma_i {M_i}^5 \over 64 \pi^4 \alpha^2
\tau^2} 
\label {PIRESTZERO}
\end{equation}
\begin{equation}
\Pi_{cont}(\tau \to 0) =
{6 {e_Q}^2 \over \pi^4 \tau^6} 
\label {PICONTZERO}
\end{equation}
One sees that as $\tau \to \infty$, the correlator is dominated
by the ground state contribution, but as $\tau \to 0$, the
continuum contribution always dominates, independent of the
mass spectrum.

We now consider the fraction $F_i$ of a given 
state $\Psi_i$ present in the $\bar{Q}Q$ correlator 
at a given Euclidian time $\tau$, 
\begin{equation}
F_i(\tau) = {\Pi_i(\tau) \over \Pi(\tau)},
\end{equation}
By definition, $F_i(\tau) \leq 1$, and for the ground state
\begin{equation}
F_0(\tau \to \infty) = 1. \label{lim}
\end{equation}
The formation time $\tau_f$ for the ground state can be defined now 
by the time at which the correlator is dominated by the ground
state contribution, or equivalently  
when the relation 
(\ref{lim}) is satisfied with a certain accuracy. 

To illustrate the 
time evolution of the contents of the $\bar{Q}Q$ correlator, 
we plot in Fig. 1 the functions $F_i(\tau)$ for $J/\Psi$ 
and for $\Upsilon$.  For the charm case, we have included the $J/\Psi$ 
and $\Psi^\prime$ resonances, the continuum starting at $M_{th} = M_D$, 
and also the next three prominent 
$\Psi(nS)$ resonances above open charm threshold.  For the
bottom quark case, we have included the $\Upsilon(1S)$, $\Upsilon(2S)$, 
and $\Upsilon(3S)$ 
resonances, the continuum starting at $M_{th} = M_B$, and again three 
prominent S-wave resonances above threshold.  To exhibit the effect of
the continuum contribution for small $\tau$, we also show curves
for these ratios with the continuum omitted.

Also shown in Fig. 1 are the discrete values of formation time 
$\tau_f^{(o)}$ which result
from the simple estimates in Eq. \ref{form12}, using the inverse mass
spacing between the ground and first excited states.  One sees that
the rapid increase in the $F_i(\tau)$ ratios occurs generally in
the region of $\tau_f^{(o)}$ values, but of course a range of formation
time values is involved in the approach towards the ground state
dominance of the correlator.  Operationally, we can then interpret
the ratio functions $F_i(\tau)$ as the generalization of the form
\begin{equation}
F_i^{(o)}(\tau) \equiv \theta(\tau - \tau_f^{(o)})
\label{SINGLEFORMATIONTIME}
\end{equation}
which would be expected if the narrow resonance state $\Psi_i$ had been 
absent until its instantaneous formation at the time $\tau_f^{(o)}$.   
We are then led to interpret the derivative of the $F_i(\tau)$ ratios
as a continuous distribution $\cal{P}(\tau)$ of formation times which occur
in the evolution of the real correlator, 
\begin{equation}
{\cal{P}}_i(\tau) = {dF_i(\tau) \over d\tau}.
\label{DIST}
\end{equation}
These distributions are shown in Fig. 2, again for the ground states of
charmonium and bottomonium, and compared with the $\delta$ - function
position at the corresponding $\tau_f^{(o)}$ values.  One sees that
the distributions peak at $\tau$ somewhat less than 
these single formation time values. The mean values for these
distributions are comparable to the $\tau_f^{(o)}$, but the widths
of the distributions are also comparable to the mean. 
Thus in evaluating
any physical quantity which depends on a resonance formation time, one
should average this quantity over the distribution in Eq. \ref{DIST}.

In principle, information is also contained in the correlator
about the time evolution and formation of the excited states of
quarkonium.  However, it is not straightforward how to extract
this information, since the procedure used for the ground states
depends on their dominance at large Euclidean times.  We have
attempted an approximate procedure, which involves omitting 
in the dispersion integral all
states with lower mass than the state under consideration. 
This of course neglects the effects of interference in the wave
function between the states under consideration and all lower
states in the wave packet evolution picture, 
or equivalently neglects all initial fluctuations with
energy less than the mass of the excited state in the virtual
state picture.  The results for $\psi^\prime$, $\Upsilon(2S)$, 
and $\Upsilon(3S)$ are shown in Fig. 3.  The general shape and
parameters are in accord with those for the ground states.  It is
interesting to note that the relatively longer formation time
for the $\psi^\prime$ is a direct result of its closeness to
the continuum threshold, whereas one might alternatively attribute this
to the larger size of the final physical state.  Of course, both of 
these pictures must be consistent with the same confinement
dynamics, and hence may be expected to be related.

In summary, we have shown how the experimental data on the current 
correlators
can be converted to the formation time distributions of the
physical states.  The necessary information in the case of 
vector heavy quarkonia is provided by the data on $e^+ e^-$ annihilation.  
The formation time distributions are nonzero in the region of time 
expected from simple arguments involving
the inverse bound state spacings, but in addition show interesting
shapes which persist to large times.  
Our results could be used as an input in phenomenological modeling of quarkonium
production 
in $p-A$ and $A-B$ collisions.

\newpage

\begin{figure}
\begin{center}
$$\psboxto(5in;0cm){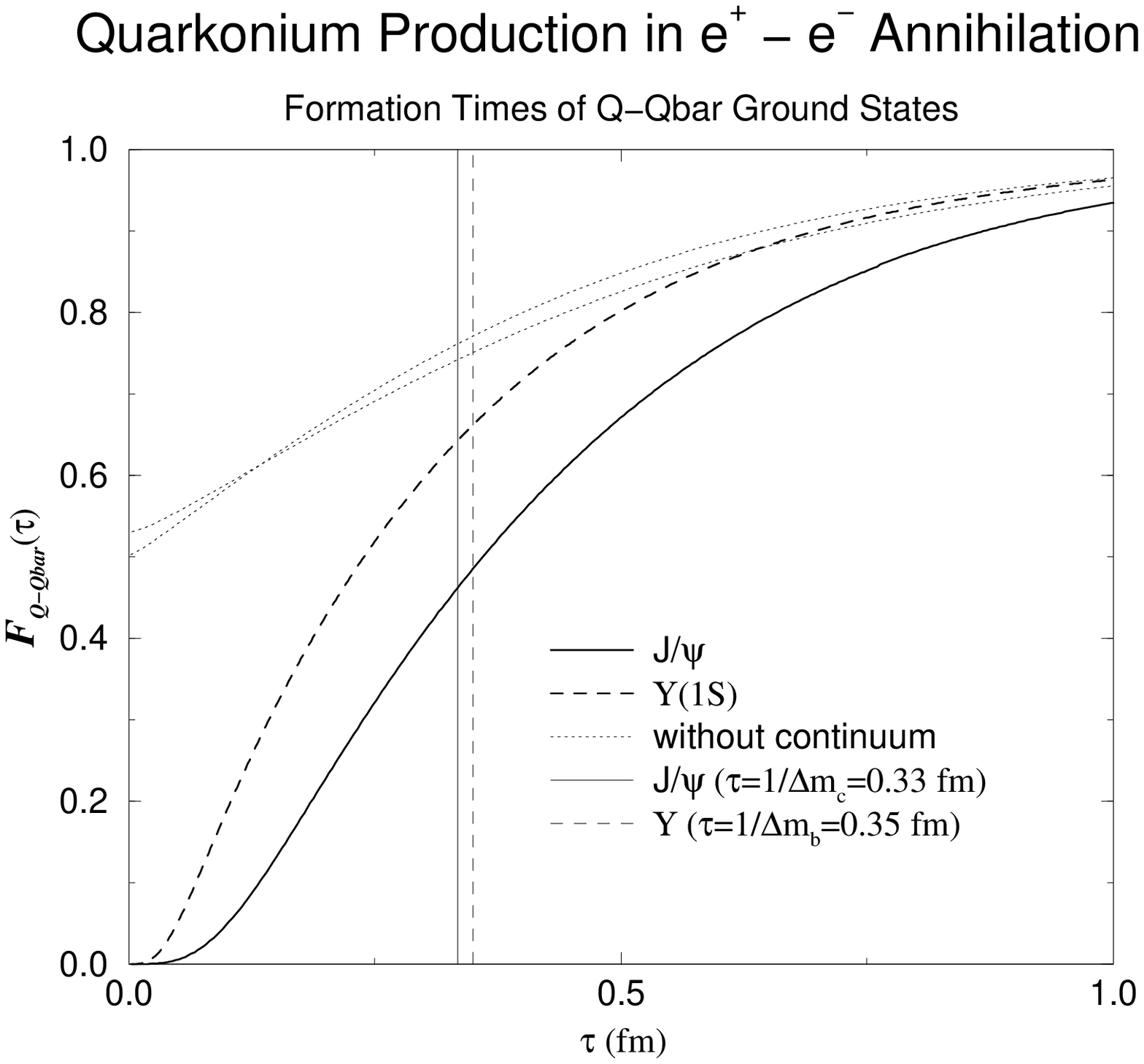}$$
\caption{ Formation times for the ground states of quarkonium
in $e^+ - e^-$ 
annihilation}
\end{center}
\end{figure}

\begin{figure}
\begin{center}
$$\psboxto(0cm;5in){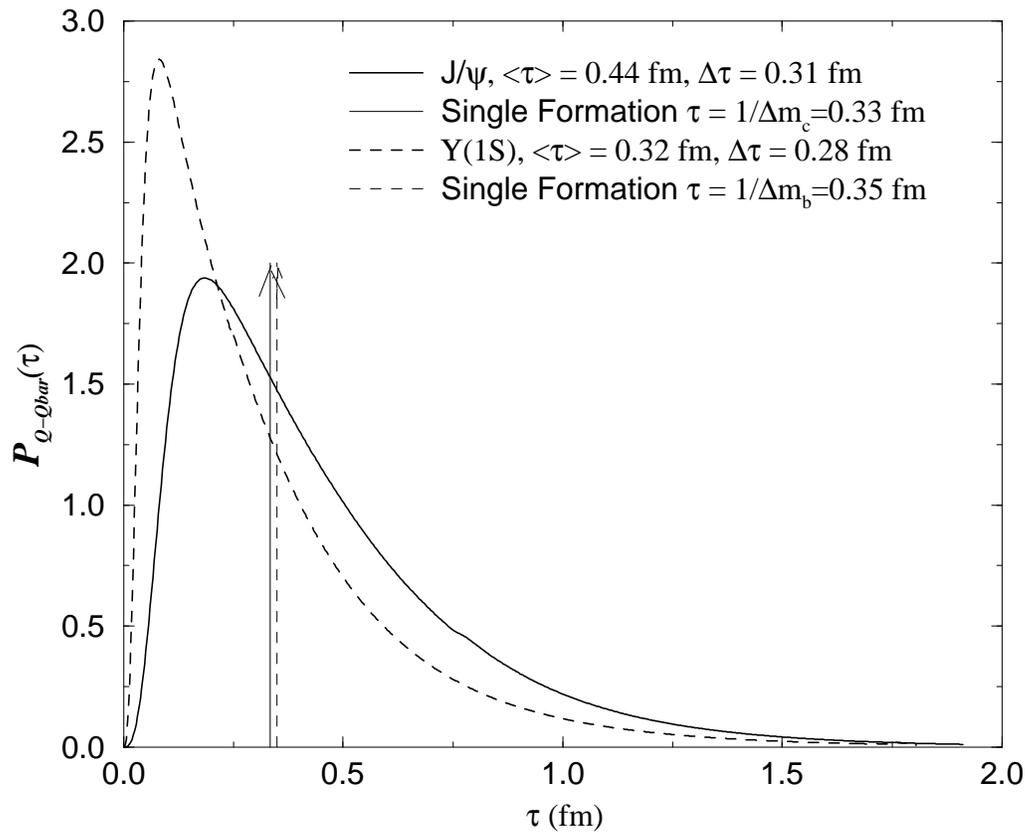}$$
\caption{ Normalized distribution functions for formation
times of $J/\psi$ and $\Upsilon$ in $e^+ - e^-$ 
annihilation}
\end{center}
\end{figure}

\begin{figure}
\begin{center}
$$\psboxto(0cm;5in){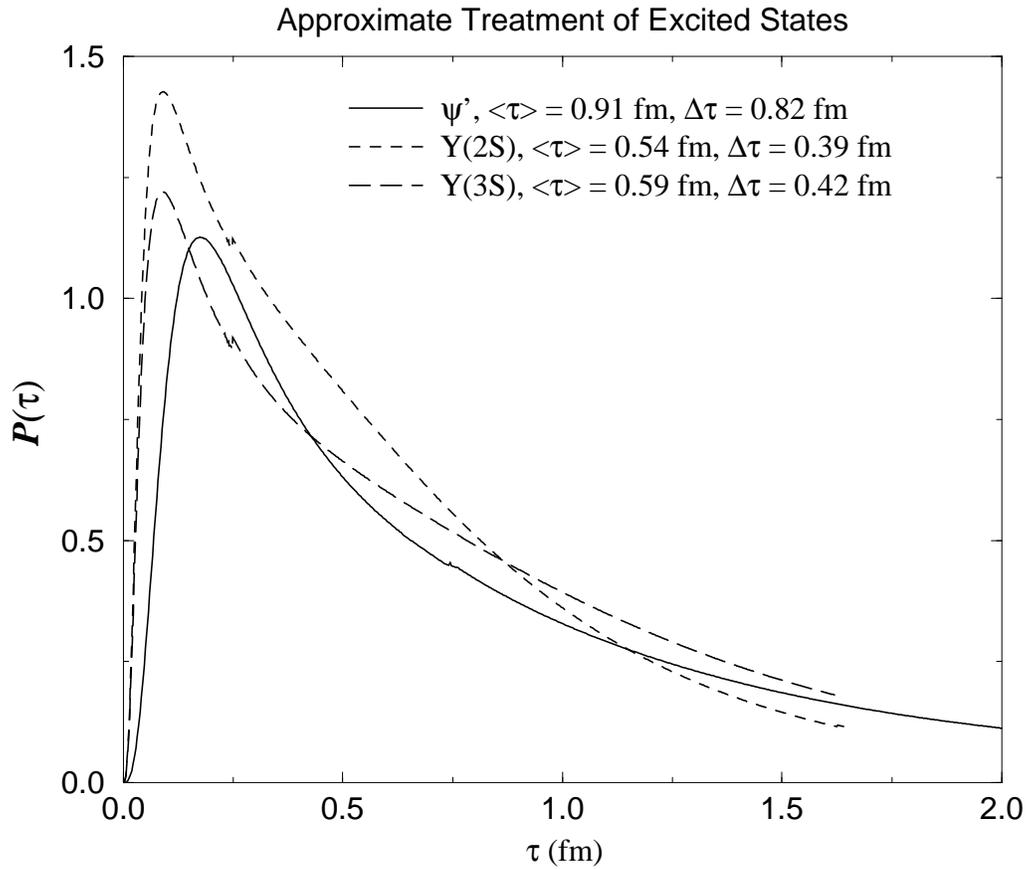}$$
\caption{ Normalized distribution functions from approximate
treatment of formation
times for excited quarkonium states in $e^+ - e^-$ 
annihilation}
\end{center}
\end{figure}
\end{document}